\documentclass[aps,prl,twocolumn,english,showpacs,supersciptaddress]{revtex4-1}
\usepackage[T1]{fontenc}
\usepackage[latin9]{inputenc}
\usepackage{graphicx}
\usepackage{amssymb}
\usepackage{color}
\usepackage{amsmath}

\usepackage{babel}
\begin{document}

\title{On the Neglect of Local Coulomb Interaction on Oxygens \\
       in Perovskites Described by the Multi-band $d-p$ Model}

\author{K. Ro\'sciszewski and A. M. Ole\'s}

\affiliation{\mbox{M. Smoluchowski Institute of Physics, Jagiellonian
University, prof. S. \L{}ojasiewicza 11, PL-30-348 Krak\'ow, Poland}}

\begin{abstract}
On the example of TiO$_4$ layer (such as realized in Sr$_2$TiO$_4$)
we study electronic structure of multi-band $d-p$ models
describing transition metal perovskites.
In agreement with the experiment, the studied system is predicted to
be a robust nonmagnetic insulator. A realistic treatment of
electronic structure requires one to introduce non-zero Coulomb local
interactions at $2p$ oxygen orbitals. However, up till now majority of
papers based upon multi-band models made an approximation of neglecting
such interactions. We show that this simplification does not lead to
serious problems in predictions of the electronic structure provided
the Coulomb interactions at titanium ions and charge transfer gap are
suitably renormalized  (so they become entirely different with respect
to the true microscopic $d-p$ model parameters).
\textit{Published in: Acta Physica Polonica A \textbf{133}, 356 (2018).}
\end{abstract}


\maketitle

\subsection{1. Introduction}

Shortly after the discovery of high temperature superconductors,
it was realized that both copper $3d$ and oxygen $2p$ orbitals
contribute to the correlated states in the cuprates and should be thus
included explicitly in the respective multi-band $d$-$p$ model for
CuO$_2$ planes \cite{Var87,Arr09} or CuO$_3$ chains \cite{Ole91}.
Nowadays a common approach is to use $d-p$ model for the description
of electronic structure of a correlated insulator for the analysis of
charge and magnetization density distribution and electron
correlations, exactly as was done in cuprates in the past
\cite{Gra92}. The multi-band $d-p$ model is richer and provides one
with a much more wide physical picture than simpler models which
feature only $d$-orbitals \cite{Mona,Bie13}. Quite recently the
multi-band $d-p$ models featuring on-site interactions defined both
on oxygens and on metal ions were investigated for ruthenium, iridium,
and titanium oxides \cite{Ros15,pss}. We focused on the full
description of Coulomb interactions including oxygen orbitals and
showed that they cannot be neglected. In contrast to this many papers
in the literature neglect such interactions on oxygens. We will show
that this simplification does not lead to serious problems in
prediction of the electronic structure provided Coulomb interactions
at titanium ions and charge-transfer-gap are properly renormalized.

In the previous paper \cite{pss} the preliminary density functional
theory computations were performed on quasi-two-dimensional
Sr$_2$TiO$_4$ and it was found that the ionicity of Sr ions is
exactly +2. Therefore for the description of the two-dimensional (2D)
model for TiO$_4$ layer one must assume that the number of valence
electrons per TiO$_4$ unit is exactly $24$.
This number is used as input to unrestricted HF computations
within  multi-band $d$-$p$ model featuring on-site
Coulomb interactions  both on titanium ions and oxygen ions.
The result is that  Sr$_2$TiO$_4$ is a robust non-magnetic insulator.
The experimental band gap of 3.8 eV was recovered from the
computations for: $U_d=9.0$ eV, $U_p=4.4$ eV and
charge-transfer gap $\Delta=\varepsilon_d - \varepsilon_p = 6.5$ eV
(defined for bare energy levels) \cite{pss}.
The $d$-electron count on Ti ion is  $n_d\simeq 1.2$,
almost equally distributed over $e_g$ and $t_{2g}$ orbitals.

Following preliminary results from ref. \cite{pss}
we provide definite  prescription how to modify
local Coulomb interaction on metal ions and charge-transfer gap
in simplified $d-p$ models which neglect local interations
on oxygens.

\subsection{2. Multi-band $d-p$ model Hamiltonian}
\label{sec:model}

The multi-band $d-p$ Hamiltonian for TiO$_4$
layer includes five $3d$ orbitals at each titanium ion and three
$2p$ orbitals at each oxygen ion. It consists of the following parts:
\begin{equation}
{\cal H}= H_{\rm kin}+ H_{\rm diag} +H_{\rm int}.
\label{model}
\end{equation}
where the different terms in Eq. (\ref{model}) stand for the kinetic
energy ($H_{\rm kin}$),
crystal-field splitting ($H_{\rm diag}$), and the intraatomic Coulomb
interactions ($H_{\rm int}$).

The kinetic (hopping) part of the Hamiltonian is:
\begin{equation}
H_{\rm kin}=   \sum_{ \{i, \mu; j, \nu\}, \sigma} t_{ i, \mu; j,\nu}
 c^{\dagger}_{i, \mu, \sigma}  c_{j, \nu, \sigma}   + H.c.,
\end{equation}
where we employ notation with $c_{j,\nu,\sigma}^{\dagger}$
standing for the creation of an electron at site $j$ in an orbital
$\nu$ with up and down spin, $\sigma=\uparrow,\downarrow$.
The model includes all $d$ orbital states on titanium,
$\nu\in\{xy,yz,zx,3z^2-r^2,x^2-y^2\}$, and three $2p$ orbitals per
oxygen atom, $\nu\in\{p_x,p_y,p_z\}$. Alternatively, i.e., choosing a
more intuitive notation, we can write $d_{j,\nu,\sigma}^{\dagger}$ for
$d$ orbitals, while $p_{j,\nu,\sigma}^{\dagger}$ for $p$ orbitals.
The matrix  elements $t_{ i, \mu; j,\nu}$ are non-zero only for
nearest-neighbor atoms; they are obtained
using Slater-Koster rules \cite{Sla54,Ros15}.

The $H_{\rm diag}$ depends only on particle operator numbers and
describes  effects of crystal field and the difference of reference
energies (i.e., charge-transfer-gap),
\begin{equation}
\label{Delta}
\Delta=\varepsilon_d-\varepsilon_p,
\end{equation}
between empty $d$ and empty $p$ orbitals.
In the following we fix reference energy at $\varepsilon_d=0$,
hence instead of $\Delta$ there is only $\varepsilon_p$ present in
$H_{\rm diag}$:
\begin{eqnarray}
H_{\rm diag}  =
\sum_{i,\mu=x,y,z;\sigma} \varepsilon_p(i)\,
p^\dagger_{i,\mu,\sigma}  p_{i,\mu,\sigma} + \\   \nonumber
\sum_{i,\mu=xy,yz,... ;\sigma} f^{cr}_{\mu,\sigma}
d^\dagger_{i,\mu,\sigma} d_{i,\mu,\sigma}.
\label{diag}
\end{eqnarray}
Note that the first sum is restricted to oxygen sites, while the second
one runs over titanium sites.

The local Coulomb interaction  for titanium ions is
\begin{eqnarray}
H_{\rm int}(d)=
  U_d \sum_{m, \mu}  n_{m, \mu, \uparrow} n_{m, \mu, \downarrow}  \nonumber \\
+\frac{1}{2}\sum_{m,\mu\neq\nu}
\left(U_d-\frac{5}{2}J_{d,\mu\nu}\right)n_{m,\mu}n_{m,\nu}    \nonumber \\
- \sum_{m,\mu\neq\nu}
J_{d,\mu\nu}\,\mathbf{S}_{m,\mu}\cdot\mathbf{S}_{m,\nu}   \nonumber \\
+  \sum_{m,\mu\neq\nu} J_{d,\mu\nu}\,
d^\dagger_{m,\mu, \uparrow} d^\dagger_{m,\mu, \downarrow}
d_{m,\nu, \downarrow}^{} d_{m,\nu, \uparrow}^{}.
\label{hubbard-intra}
\end{eqnarray}
where again $\mu, \nu$  enumerate $d$-orbitals, and
$J_{d,\mu\nu}$ is the tensor of on-site interorbital exchange
elements  \cite{Ole2005}).
In our computations the whole anisotropic tensor $J_{d,\mu\nu}$ is
treated without any simplifications. The formula for
local Coulomb interactions at oxygen sites (for $2p$ orbitals) is
analogous.

The effective $d-p$ model  requires a number
of parameters. The  in-plane hopping elements were fixed as
$(pd\sigma)=-2.4 $ eV and $(pd\pi) =1.3$ eV and also
$(pp\sigma)=0.6$ eV and $(pp\pi)=-0.15$ eV
\cite{Miz96,Moc04,Wad09}.
The choice of the Coulomb elements in $d-p$ model is difficult.
There are reliable estimates for $U_d$ ($\sim   4$ eV)
but only in effective models featuring solely composite $d$-type
Wannier orbitals (i.e.,  explicit treatment of $p$-orbitals is absent).
The typical parameter $U_d\sim 4$ eV (for titanium ions)
is different from that which should be used in the framework of the
multi-band $d-p$ model.  Namely, it is smaller
in the $d$-orbital-only model from that used in the $d-p$ model by
$\sim$50 \% due to a massive screening.

\begin{figure}[t!]
\centerline{\includegraphics[width=\columnwidth]{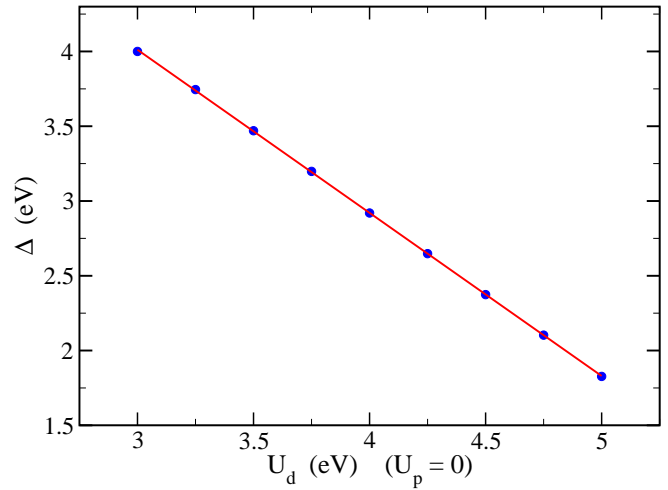}}
\caption{
Charge-transfer energy $\Delta$ (\ref{Delta}) versus $U_d$ for the
parameters which reproduce experimental band gap of 3.8 eV.
The extrapolated dependence is linear ---
$\Delta \approx 7.28 - 1.09\,\, U_d$ (with $U_d$ in eV).
Fixed parameters of the {\it d-p} Hamiltonian are:
$U_p=0$, $J_0=0$; crystal field $t_{2g}$ to $e_g$ splitting is 2.0 eV
and $J_d =0.8$ eV.
 }
\label{fig1}
\end{figure}

We decided to study several possibilities, namely
$U_d=4$, 6, 8, 9, and 10  eV according to the data in the literature:
\cite{Spr03} ($U_d\in [7,8]$ eV); \cite{ong2011,Par10,Par12}
($U_d\sim 8$ eV); and $U_d\sim 6$ eV \cite{Ani91,Hav12}).
Hund's exchange elements are less screened than intraorbital Coulomb
elements and are closer to their atomic values. For Hund's exchange
$J_d$ (between two $t_{2g}$ electrons) and for our computations we
decided to select $J_d=0.8$ eV according to refs. \cite{Ani91,Str05}
(also other possibilities were studied, namely $J_d=0.9$ eV and
$J_d=1.0$ eV). The Coulomb repulsion $U_p$ for oxygen $p$-orbitals
was fixed as $U_p=4.4$ eV like in refs. \cite{Hyb89,Woh13,Ros15}.
while Hund's exchange was fixed as $J_p=0.8$ eV  \cite{Str05}.
Next the  average $t_{2g}$ to $e_g$
crystal-field splitting (for bare levels) which we studied were
either 2.0 eV or 3.0 eV \cite{Ike09,zhou2011,Moc04}.
Finally, very important parameter for the charge distribution in
(TiO$_2)^{2-}$ planes is a charge-transfer gap $\Delta$ (\ref{Delta})
which was examined as a continuous parameter in the range from 1.0
eV up to 9.0 eV.

We used the unrestricted HF approximation to investigate the
TiO$_4$ cluster ($4\times 4$ TiO$_4$ units with cyclic boundary
conditions) and with $N_e=24$ electrons per single TiO$_4$ unit. The
implementation of the model Hamiltonian (\ref{model}) is the same as
described in refs. \cite{Miz96,Miz01,Ros15,Ave13}.
The possibilities of different types of order were taken into account:
nonmagnetic phase, ferromagnetic and antiferromagnetic
with spins aligned along: (1,1,0) or (1,0,0) or (0,0,1) direction.
We performed runs for all combinations of the Hamiltonian
parameters. Any reasonable parameter set leads to nonmagnetic
insulator as the ground state. The experimental band gap in
Sr$_2$TiO$_4$ is 3.8 eV \cite{Mat05} and it is reproduced when the
Hamiltonian parameters are properly tuned to $U_d=9.0$ eV, $J_d=0.8$ eV
and $\Delta\simeq 6.5$ eV \cite{pss}.

\subsection{3. Consequences of the neglect of Coulomb repulsion \\
on oxygens}
\label{sec:noUp}

Let us now study what happens when Coulomb repulsion on oxygens is
neglected. We argue that in general it is not correct to set
$U_p=J_p=0$ but this approximation/simplification was used by almost
all the authors up to now.

Thus, we performed multiple HF computations using  Hamiltonian
parameter sets with $U_p = J_p = 0$ and with numerous differently
renormalized $U_d$ and $\Delta$.
Much to our surprise we obtained quite normal electronic structure
with  nonmagnetic ground state  as a generic solution.
When studying various possibilities how to renormalize
$U_d$  and $\Delta$  so as the obtained results
are physically sensible  we obviously should look after
such sets of  $U_d$ and $\Delta$ which reproduce
experimental band gap  of 3.8 eV.
The results coming out from this approach are shown in Fig.~1.

\subsection{4. Renormalization of $U_d$ and $\Delta$ for $U_p=0$ models}

From Fig. 1 if follows  that if one decides to model the titanium
perovskite neglecting the local Coulomb repulsion on oxygen
($U_p=J_p=0$) and in addition if one accepts $U_d= 4$ eV then
the charge-transfer energy should be chosen as $\Delta \approx 3.0$ eV.
Note that the usual parameters of the $d-p$ multi-band model are by
$~50$\% larger, just to remind: $U_d= 9$ eV and $\Delta = 6.5$ eV.

If a different  value of  $U_d$   is advocated in a particular paper
one can pick out from the literature
still we can predict the precise value of $\Delta$ which should be
appropriate in such the context.

\subsection{5. Summary}

We have taken the number of electrons which follows from the ionic
configuration of Sr$^{2+}$ ions for a periodic $4\times 4$ TiO$_4$
cluster, as suggested by \textit{ab initio} electronic structure
calculations and examined in electronic distribution within
unrestricted HF computations performed using the multi-band $d$-$p$
model. Several possibilities for Hamiltonian parameters were studied.
For any reasonable choice of the parameters we find a good nonmagnetic
insulator, with a gap reproduced for $U_d=9.0$ eV, $U_p=4.4$ eV,
and a large charge-transfer gap $\Delta=6.5$~eV.

We also considered a simplified set of parameters with $U_p=J_p=0$
and find that this choice implies considerable reduction of the
remaining parameters:
($i$) the value of $U_d$ is smaller by $\sim$ 50 \%;
($ii$) the charge-transfer gap $\Delta$  is also reduced to $\sim 3.0$
eV. Only taking these renormalization one can reproduce the essential
features of the electronic structure of Sr$_2$TiO$_4$. We remark that
while such a renormalization is possible for the bulk, the treatment
of systems with non-equivalent oxygen and transition metal positions
would require a complete parameter set with finite electron
interactions at oxygen $2p$ orbitals.

\subsection{Acknowledgments}
\vskip -.3cm

We kindly acknowledge support by Narodowe Centrum Nauki
(NCN) under Project No. 2012/04/A/ST3/00331.



\vskip -.3cm
\begin{thebibliography}{99}

\bibitem{Var87} V.J.~Emery,
                   \textit{Phys. Rev. Lett.} \textbf{58}, 2794 (1987);
                C.M.~Varma, S.S.~Schmitt-Rink, E.E.~Abrahams,
                   \textit{Solid State Commun.} \textbf{62}, 681 (1987); 
                A.M.~Ole\'s, J.~Zaanen, P.~Fulde,
                   \textit{Physica B\&C} \textbf{148}, 260  (1987).

\bibitem{Arr09} E.~Arrigoni, M.~Aichhorn, M.~Daghofer, W.~Hanke,
                   \textit{New J. Phys.} \textbf{11}, 055066 (2009).

\bibitem{Ole91} A.M. Ole\'s, W. Grzelka,
                   \textit{Phys. Rev. B} \textbf{44}, 9531 (1991).

\bibitem{Gra92} J.B.~Grant, A.K.~McMahan,
                   \textit{Phys. Rev. B} \textbf{46}, 8440 (1992).

\bibitem{Mona}  M.~M\"oller, G.A.~Sawatzky, M.~Berciu,
                   \textit{Phys. Rev. Lett.} \textbf{108}, 216403 (2012);
                   \textit{Phys. Rev. B} \textbf{86}, 075128 (2012).

\bibitem{Bie13} K. Bieniasz, A.M.~Ole\'s,
                   \textit{Phys. Rev. B} \textbf{88}, 115132 (2013).

\bibitem{Ros15} K.~Ro\'sciszewski, A.M.~Ole\'s,
                   \textit{Phys. Rev. B} \textbf{91}, 155137 (2015); \\
                                         \textbf{93}, 085106 (2016).

\bibitem{pss}   K.~Ro\'sciszewski, P. Piekarz, A.M.~Ole\'s,
                   \textit{Physica Status Solidi (b)} \textbf{254}, 1700022 (2017).

\bibitem{Sla54} C.~Slater, G.F.~Koster,
                   \textit{Phys. Rev.} \textbf{94}, 1498 (1954).

\bibitem{Ole2005} A.M.~Ole\'s, G.~Khaliullin, P.~Horsch, L.F.~Feiner,
                   \textit{Phys. Rev. B\/} \textbf{72}, 214431 (2005).

\bibitem{Miz96} T.~Mizokawa, A.~Fujimori,
                   Phys. Rev. B \textbf{54}, 5368 (1996).

\bibitem{Moc04} M.~Mochizuki,
                   J. Phys. Soc. Jpn. \textbf{71}, 2039 (2002);   
                M.~Mochizuki, M.~Imada,
                   J. Phys. Soc. Jpn. \textbf{73}, 1833 (2004);
                   New~J.~Phys. \textbf{6}, 154 (2004).

\bibitem{Wad09} H.~Wadati, A.~Chikamatsu, M.~Takizawa, H.~Kumigashira, T.~Yoshida,
                   T.~Mizokawa, A.~Fujimori, M.~Oshima, N.~Hamada,
                   J. Phys. Soc. Jpn. \textbf{78}, 094709 (2009).

\bibitem{Spr03} A.I.~Lichtenstein, V.I.~Anisimov, M.I.~Katsnelson,
                   {\it Electronic structure of transition metal oxides},
                   in: Electronic Structure and Magnetism of Complex Materials,
                   edited by D.J.~Singh and D.A.~Papaconstantopoulos
                   (Springer-Verlag, Berlin, Heidelberg,2003)

\bibitem{ong2011} P.V.~Ong, J.~Lee, W.E.~Pickett,
                     Phys. Rev. B \textbf{83}, 193106 (2011).

\bibitem{Par10} S.G.~Park, B.~Magyari-K\"{o}pe, Y.~Nishi,
                   Phys. Rev. B \textbf{82}, 115109 (2010).

\bibitem{Par12} B.~Magyari-K\"{o}pe, S.G.~Park, H.~Lee, Y.~Nishi,
                   J.~Mater. Sci. \textbf{47}, 7498  (2012).

\bibitem{Hav12} M.W.~Haverkort, M.~Zwierzycki, O.K.~Andersen,
                   Phys. Rev. B \textbf{85}, 165113 (2012).

\bibitem{Ani91} V.I.~Anisimov, J.~Zaanen, O.K.~Andersen,
                   Phys. Rev.B \textbf{44}, 943 (1991).

\bibitem{Str05} S.V.~Streltsov, A.S.~Mylnikova, A.O.~Shorikov,
                   Z.V.~Pchelkina, D.I.~Khomskii, V.I.~Anisimov,
                   Phys. Rev. B \textbf{71}, 245114 (2005).

\bibitem{Hyb89} M.S.~Hybertsen, M.~Schl\"uter, N.E.~Christensen,
                   Phys. Rev. B \textbf{39}, 9028 (1989).
%

\bibitem{Woh13} K.~Wohlfeld, S.~Nishimoto, M.W.~Haverkort, J.~van den Brink,
                   Phys. Rev. B \textbf{88}, 195138 (2013).

\bibitem{Ike09} H.~Ikeno, F.M.F.~de Groot, E.~Stavitski, I.~Tanaka,
                   J.~Phys.: Condens. Matter \textbf{21}, 104208 (2009);\\
                H.~Ikeno, H.~Mizoguchi, I.~Tanaka,
                   Phys. Rev. B \textbf{83}, 155107 (2011).
%

\bibitem{zhou2011} Z. Ke-Jin, M.~Radovic, J.~Schlappa, V.~Strocov, R. Frison,
                      J.~Mesot, K.~Patthey, T.~Schmitt,
                      Phys. Rev. B \textbf{83}, 201402(R) (2011).

\bibitem{Miz01} T.~Mizokawa, L.H.~Tjeng, G.A.~Sawatzky, G.~Ghiringhelli,
                     O.~Tjernberg, N.B.~Brookes, H.~Fukazawa, S.~Nakatsuji,
                     Y.~Maeno,
                     Phys. Rev. Lett. \textbf{87}, 077202 (2001).

\bibitem{Ave13} A.~Avella, P.~Horsch, A.M.~Ole\'s,
                   Phys. Rev. B \textbf{87}, 045132 (2013);
                A.~Avella, A.M.~Ole\'s, P.~Horsch,
                   Phys. Rev. Lett. \textbf{115}, 206403 (2015).

\bibitem{Mat05} J.~Matsuno, Y.~Okimoto, M.~Kawasaki, Y.~Tokura,
                   Phys. Rev. Lett. \textbf{95}, 176404 (2005).


\end{thebibliography}
\end{document}